\documentclass[journal]{IEEEtran}

\usepackage{cite}
\usepackage{hyperref}

\ifCLASSINFOpdf
    \usepackage[pdftex]{graphicx}
    \DeclareGraphicsExtensions{.pdf,.jpeg,.png}
\else
  \usepackage[dvips]{graphicx}
\fi
\usepackage{amsmath}
\usepackage{amssymb}
\usepackage{gensymb}
\ifCLASSOPTIONcompsoc
  \usepackage[caption=false,font=normalsize,labelfont=sf,textfont=sf]{subfig}
\else
  \usepackage[caption=false,font=footnotesize]{subfig}
\fi
\usepackage{url}

    \def\Complex{{\rm\rule[.23ex]{.03em}{1.1ex}\kern-.3em{C}}}

    \newcommand{\be}{\begin{equation}} \newcommand{\ee}{\end{equation}}
    \newcommand{\bea}{\begin{eqnarray}} \newcommand{\eea}{\end{eqnarray}}
    \newcommand{\benum}{\begin{enumerate}} \newcommand{\eenum}{\end{enumerate}}

    \newcommand{\qa}{{\bf a}}

    \newcommand{\qB}{{\bf B}}

    \newcommand{\qH}{{\bf H}}

    \newcommand{\qP}{{\bf P}}

    \newcommand{\bbC}{{\mathbb C}}

    \usepackage{xcolor}

    \newcommand{\rl}[1]{\color{red}#1}

\begin{document}
%
\title{Geo2ComMap: Deep Learning-Based MIMO Throughput Prediction Using Geographic Data}
%
%
%

\author{Fan-Hao~Lin,~Tzu-Hao~Huang,~Chao-Kai~Wen,~\IEEEmembership{Fellow,~IEEE},~Trung~Q.~Duong,~\IEEEmembership{Fellow,~IEEE}

\thanks{F.-H. Lin, T.-H. Huang, and C.-K. Wen are with the Institute of Communications Engineering, National Sun Yat-sen University, Kaohsiung 80424, Taiwan, Email: \{22135lin, peter94135\}@gmail.com, chaokai.wen@mail.nsysu.edu.tw.}

\thanks{T. Q. Duong is with Memorial University, Canada, and also with Queen's University Belfast, UK, Email: tduong@mun.ca.}
}

\maketitle
\begin{abstract}
Accurate communication performance prediction is crucial for wireless applications such as network deployment and resource management. Unlike conventional systems with a single transmit and receive antenna, throughput (Tput) estimation in antenna array-based multiple-output multiple-input (MIMO) systems is computationally intensive, i.e., requiring analysis of channel matrices, rank conditions, and spatial channel quality. These calculations impose significant computational and time burdens. This paper introduces Geo2ComMap, a deep learning-based framework that leverages geographic databases to efficiently estimate multiple communication metrics across an entire area in MIMO systems using only sparse measurements. To mitigate extreme prediction errors, we propose a sparse sampling strategy. Extensive evaluations demonstrate that Geo2ComMap accurately predicts full-area communication metrics, achieving a median absolute error of 27.35 Mbps for Tput values ranging from 0 to 1900 Mbps.
\end{abstract}


%
\IEEEpeerreviewmaketitle

\section{Introduction}
%
%
%
%

\IEEEPARstart{R}{adiomaps} play a crucial role in future intelligent communication systems, as they enable the integration of environmental knowledge into system design to support high-throughput and reliable communication. High-precision radiomaps are essential for applications such as cellular network planning, fingerprint-based localization, and UAV path planning \cite{Zhang-24ComMag,Zhou-25ArXiv, elsayed2024ofdm}. Deterministic simulations using propagation models, such as ray tracing, are widely used for radiomap generation.
However, the high computational complexity of ray tracing limits its practicality. Recent advances in deep learning (DL) and the availability of large-scale datasets have facilitated DL-based radiomap estimation \cite{Levie-21TW,Zhou-25ArXiv,Li-SySPAN24} (see \cite{Yapar-24} for a survey), achieving high accuracy while significantly reducing computational costs and processing time. Notably, \cite{Li-SySPAN24} introduced Geo2SigMap, a framework that reconstructs dense path gain maps using geographic data and sparse site measurements with excellent accuracy.

Despite these advancements, path gain alone is insufficient for comprehensive wireless communication analysis. In conventional systems with a single transmit and receive antenna, path gain directly correlates with throughput (Tput). However, in antenna array-based MIMO systems---prevalent in 5G and emerging 6G networks---Tput estimation requires modeling spatial signal interactions, channel matrices, and key indicators such as the Rank Indicator (RI) and Channel Quality Indicator (CQI) \cite{3gpp_38_306}. Consequently, Tput estimation is significantly more resource-intensive than path gain mapping. Building on the foundation of Geo2SigMap \cite{Li-SySPAN24}, this paper introduces Geo2ComMap, a comprehensive framework that leverages geographic databases to estimate multiple communication metrics in MIMO-OFDM systems. Our key contributions include:

{\bf Dataset Generation:} Traditional radiomaps do not include datasets linking geographic maps to MIMO-OFDM Tput, posing a challenge for deep learning-based estimation. To address this, we integrate a MIMO-OFDM Tput simulation program with geographic databases \cite{Li-SySPAN24} to generate a comprehensive training dataset, enabling effective DL model development.

{\bf Deep Learning Model:} We propose U-Net-TP, the first U-Net-based architecture designed to predict multiple communication metric maps using geographic data and sparse measurements. U-Net-TP generalizes across different locations without requiring retraining. We extensively evaluate its performance across various input configurations, providing key insights into geographic-to-communication mappings.

{\bf Sampling Selection Strategy:} To mitigate high-error regions, we introduce a sparse map design strategy that reduces extreme prediction errors without increasing system costs.

\section{System Design}
\label{System Design}

This section outlines the methodology of {\bf Geo2ComMap}, which comprises three primary modules. The first module converts a geographical Building Map into corresponding channel matrices. The second module computes system Tput based on these channel matrices. The final module leverages a deep learning method to enable efficient mapping from Building Map to communication metrics.

\subsection{Building Map to Channel Matrix}
\label{Building Map to Channel Matrix}
To build a comprehensive training dataset for the deep learning model, we adopt a systematic approach inspired by \cite{Li-SySPAN24}. This involves generating a Building Map and conducting ray-tracing simulations.

For a given geographical area of dimensions $L_x \times L_y$ (in square meters), we first generate a Building Map, $\qB \in \mathbb{R}^{N_x \times N_y}$, with a resolution of $r$, where ${L_x}/{N_x} = {L_y}/{N_y} = r$ meters. This map is accompanied by a 3D mesh model, generated using:
\begin{itemize}
    \item {\bf OpenStreetMap (OSM)} \cite{openstreetmap}, an open-source geographic database, to extract building data from a selected area.
    \item {\bf Blender} \cite{blender}, an open-source 3D graphics tool, to convert the extracted data into detailed 3D building models.
\end{itemize}
Using these 3D models, we conduct ray-tracing simulations with {\bf Sionna} \cite{sionna_rt} to compute key propagation characteristics. The simulations consider both isotropic and directional antennas operating at a carrier frequency of 3.5 GHz. The isotropic antenna has a gain of 0 dBi, while the directional antenna has a boresight gain of 6.3 dBi and a horizontal/vertical half-power beamwidth of $65^{\circ}/8^{\circ}$, typical for cellular antennas. The directional antenna's orientation is randomly set within the azimuth range of $[0, 2\pi)$.

The base station (BS), serving as the transmitter, is randomly positioned within the x-y plane at a height of 5 meters above the tallest structure in the area. The user equipment (UE), acting as the receiver, is uniformly distributed across outdoor locations at a height of 1.5 meters. Using Sionna, we generate a Path Gain (PG) Map, $\qP_{\rm iso} \in \mathbb{R}^{N_x \times N_y}$, for each area, assuming an isotropic antenna at the BS, with values ranging from -160 dBm to 0 dBm. Alternatively, the PG Map can be estimated using DL-based methods such as Geo2SigMap \cite{Li-SySPAN24}.

While PG Maps can be efficiently generated without detailed system parameters, calculating MIMO system Tput requires a joint consideration of channel propagation characteristics and system-specific factors. Key propagation characteristics include channel coefficients $\alpha_l$, propagation delay $\tau_l$, and the zenith/azimuth angles of departure (AoD) $(\phi_l^\text{t}, \theta_l^\text{t})$ and arrival (AoA) $(\phi_l^\text{r}, \theta_l^\text{r})$. Combining these elements within a MIMO-OFDM system with $N_\text{t}$ transmit and $N_\text{r}$ receive antennas, the channel matrix at subcarrier $n$ is expressed as:
\begin{equation} \label{eq_H}
    \qH_n = \sum_{l=1}^{L} \alpha_{l} e^{-j 2\pi f_n \tau_l} \qa_\text{r}(\phi_l^\text{r}, \theta_l^\text{r}) \qa_\text{t}^{H}(\phi_l^\text{t}, \theta_l^\text{t}),
\end{equation}
where $L$ is the number of multipath components, $f_n$ is the corresponding subcarrier frequency, and $\qa_\text{r}(\cdot) \in \bbC^{N_\text{r} \times 1}$ and $\qa_\text{t}(\cdot) \in \bbC^{N_\text{t} \times 1}$ are the array response vectors of the UE and BS, respectively.

Using ray-tracing simulations, \eqref{eq_H} is applied to generate channel matrices for both isotropic and directional antennas. The differences between the two antenna types are inherently captured through the ray-tracing process, which accounts for variations in channel coefficients.

\subsection{Channel Matrix to System Tput}
\label{MIMO Tput Simulation}


Tput estimation involves evaluating the RI and CQI, which determine the number of spatial streams and the modulation and coding scheme (MCS), respectively. According to 3GPP TS 38.214 \cite{3gpp_38_214}, CQI levels range from 1 to 15, each corresponding to a specific spectral efficiency.
The process of determining RI and CQI follows these steps:
\begin{enumerate}
    \item Compute the autocorrelation matrix over a group of subcarriers from the channel matrices in \eqref{eq_H}, and perform eigenvalue decomposition to obtain spatial precoders and their corresponding equivalent channel gains.

    \item Evaluate performance metrics for all possible RI and CQI combinations based on the equivalent channel gains.

    \item Select the RI and CQI combination that maximizes spectral efficiency under current channel conditions.
\end{enumerate}
Once the RI and CQI are determined, the system Tput is calculated according to 3GPP TS 38.306 \cite{3gpp_38_306} as:
\begin{equation} \label{eq_TP}
\text{TP} = 10^{-6} \cdot v \cdot \log_2 Q \cdot R \cdot \frac{N_\text{RE}}{T_s},~~\text{(Mbps)}
\end{equation}
where ${1 \leq v \leq \min(N_\text{t}, N_\text{r})}$ represents the RI; $Q$ and $R$ denote the MCS associated with the CQI;  ${N_\text{RE} = 135 \times 12 = 1620}$ represents the number of resource elements per OFDM symbol; $T_s = 10^{-3}$ seconds is the duration of one OFDM symbol; the factor $10^{-6}$ converts the unit from bps to Mbps.
The resulting Tput Map, $\mathbf{TP} \in \mathbb{R}^{N_x \times N_y}$, represents the spatial distribution of Tputvalues, ranging from 0 to 1900 Mbps, across different locations within the Building Map, $\qB$.

\begin{figure}[t]
    \centering
    \includegraphics[width=3.5in]{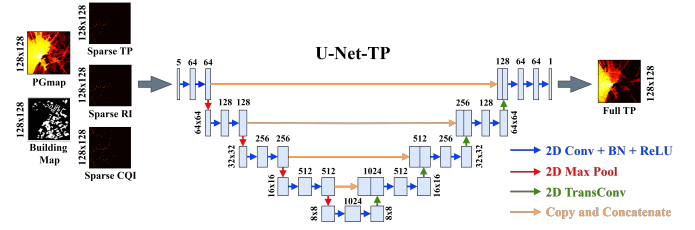} 
    \caption{Architecture of Geo2ComMap, which is a 5-input channel U-Net designed to achieve efficient and precise Tput prediction.}
    \label{fig:unet_TP}
\end{figure}

\subsection{Deep Learning Model for Tput Prediction}
\label{U-Net for Tput Prediction}

The previous subsections have demonstrated that generating a Tput Map involves a multi-step process. Each step requires complex computations and substantial processing time. To address this challenge, the core of {\bf Geo2ComMap} is a deep learning model designed to efficiently predict a complete Tput Map for a given area using a Building Map and sparse Tput-related measurements.
The proposed model, named {\bf U-Net-TP}, is illustrated in Fig.~\ref{fig:unet_TP}. U-Net-TP takes a 5-channel input image:
$[\qB, \qP_{\rm iso}, \mathbf{RI}_{\rm s}, \mathbf{CQI}_{\rm s}, \mathbf{TP}_{\rm s}] \in \mathbb{R}^{N_x \times N_y \times 5}$, 
where $\qB$ represents the Building Map, $\qP_{\rm iso}$ is the isotropic PG Map, and $\mathbf{RI}_{\rm s}, \mathbf{CQI}_{\rm s}, \mathbf{TP}_{\rm s}$ are sparse maps containing RI, CQI, and Tput measurements from a limited number of UE locations.  The model outputs a complete Tput Map, $\mathbf{TP}$.

U-Net-TP consists of nine convolutional blocks and four downsampling/upsampling layers. Each convolutional block includes two 2D convolutional layers with a kernel size of $3 \times 3$, followed by batch normalization and ReLU activation. Downsampling is achieved through $2 \times 2$ max pooling, while upsampling is performed via 2D transposed convolution layers.
In the contracting path, the number of channels doubles from 64 to 1024, while the spatial dimensions are halved at each stage. Skip connections link corresponding layers in the contracting and expanding paths, helping to retain spatial information and mitigate gradient vanishing or explosion during training.

Notably, U-Net-TP shares the same underlying structure as the model used for PG Map generation \cite{Li-SySPAN24}, which is a simpler task. However, the key differences lie in the input and output feature sets. In the next section, we will discuss the relevance of input features and the feasibility of using U-Net-TP to generate both Tput and PG Maps simultaneously.

\begin{figure}[t]
    \centering
    \subfloat[\label{Building Map 1}]{\includegraphics[width=0.45\linewidth]{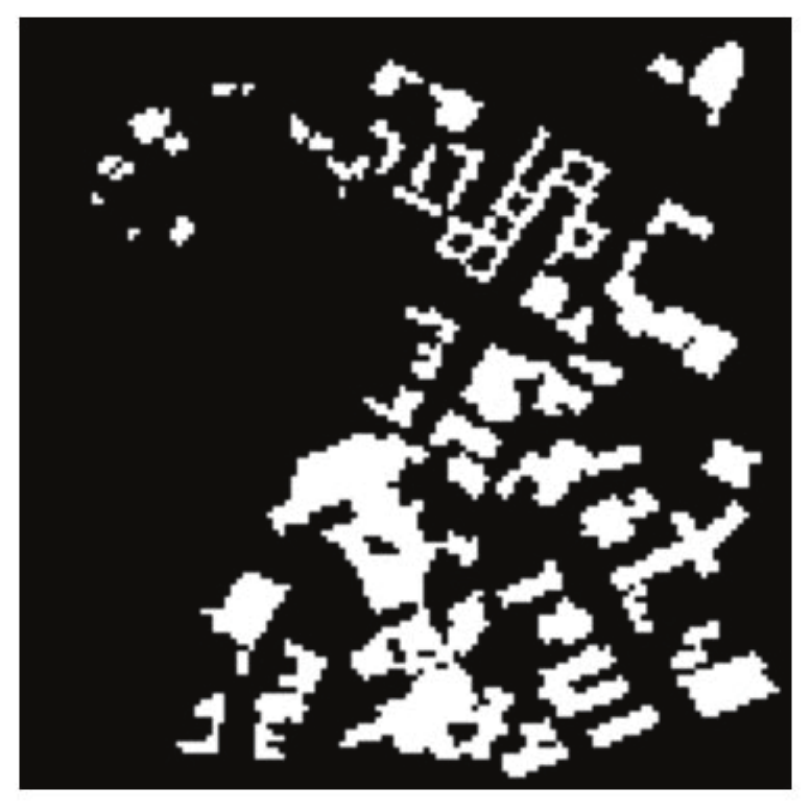}}
    \hspace{0.05\linewidth}
    \subfloat[\label{Building Map 2}]{\includegraphics[width=0.45\linewidth]{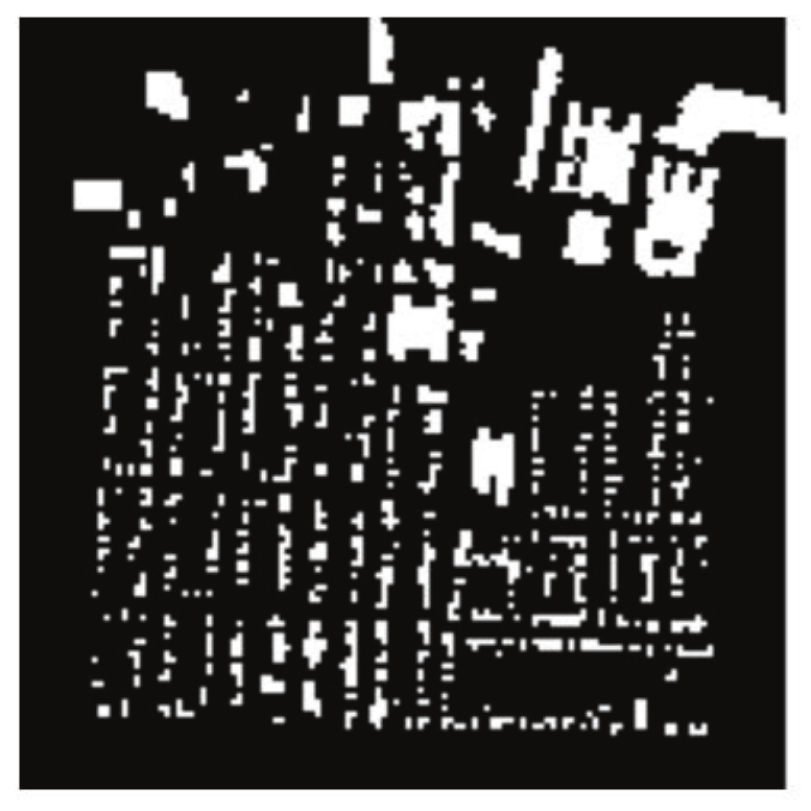}}
    \caption{Building Maps used in the experiments. (a) BM1, characterized by a prominent open space. (b) BM2, featuring densely clustered buildings.}
    \label{fig:testing_buildings}
\end{figure}

\section{Discussions}
\subsection{Dataset Generation and U-Net Training} \label{Dataset Generation and U-Net Training}

To train U-Net-TP in Geo2ComMap, we generate Building Maps and ray tracing datasets following the procedure described in Section \ref{Building Map to Channel Matrix}. The dataset consists of 32 distinct, non-overlapping areas across North America. Data generation is conducted using Sionna 0.15.1 and Blender 3.6. For each area, 100 BS locations are simulated with five antenna power levels ranging from 40 to 50 dBm, yielding a total of 15,920 datasets (excluding erroneous data).
To generate complete ground truth communication-metric maps, we consider a MIMO-OFDM system with $N_\text{t}=N_\text{r} = 4$ and a bandwidth of 100 MHz. The channel matrix $\qH$ is derived from propagation paths using \eqref{eq_H}. A constant noise level of $-93.98 , \mathrm{dBm}$ is assumed in all cases.
To generate sparse RI, CQI, and Tput maps, we sample 200, 300, 400, 500, and 600 points from the full RI, CQI, and Tput Maps obtained through the MIMO Tput simulation described in Section \ref{MIMO Tput Simulation}. These correspond to 1.22\%, 1.83\%, 2.44\%, 3.05\%, and 3.66\% of the total area, respectively. Detailed analyses of the impact of varying sampling densities and distributions are provided in the subsequent subsections.

The dataset is divided into training, validation, and testing sets with an 84\%, 10\%, and 6\% split, respectively. To ensure robust evaluation, this division is performed at the Building Map level, ensuring no overlap between the Building Maps used for training, validation, and testing. Data augmentation techniques---such as rotations (0\degree, 90\degree, 180\degree, and 270\degree) and mirroring of both input and output maps---are applied, expanding the dataset size fivefold. The model is trained using the mean squared error (MSE) loss function on the target communication metrics for 50 epochs on an NVIDIA GeForce GTX 1080 Ti GPU. For testing, 1,000 data samples are utilized, equally distributed across two scenarios: 500 samples for Building Map 1 (BM1) and 500 samples for Building Map 2 (BM2), as shown in Fig.~\ref{fig:testing_buildings}. BM1 features a large open space, whereas BM2 consists of densely clustered buildings.

\subsection{Effect of RI and CQI Information}
\label{Effect of RI and CQI information}

The Tput in \eqref{eq_TP} is highly dependent on RI and CQI values. Therefore, it is reasonable to use sparse RI and CQI maps, $(\mathbf{RI}_{\rm s}, \mathbf{CQI}_{\rm s})$, as inputs for U-Net-TP. To assess the impact of these input features on Tput prediction, we compared two U-Net-TP architectures: one utilizing a 5-channel input image, as described in Sec.~\ref{U-Net for Tput Prediction}, and another with a 3-channel input image, given by $[\qB, \qP_{\rm iso}, \mathbf{TP}_{\rm s}] \in \mathbb{R}^{N_x \times N_y \times 3}$.

\begin{figure}[t]
    \centering
    \includegraphics[width=\linewidth]{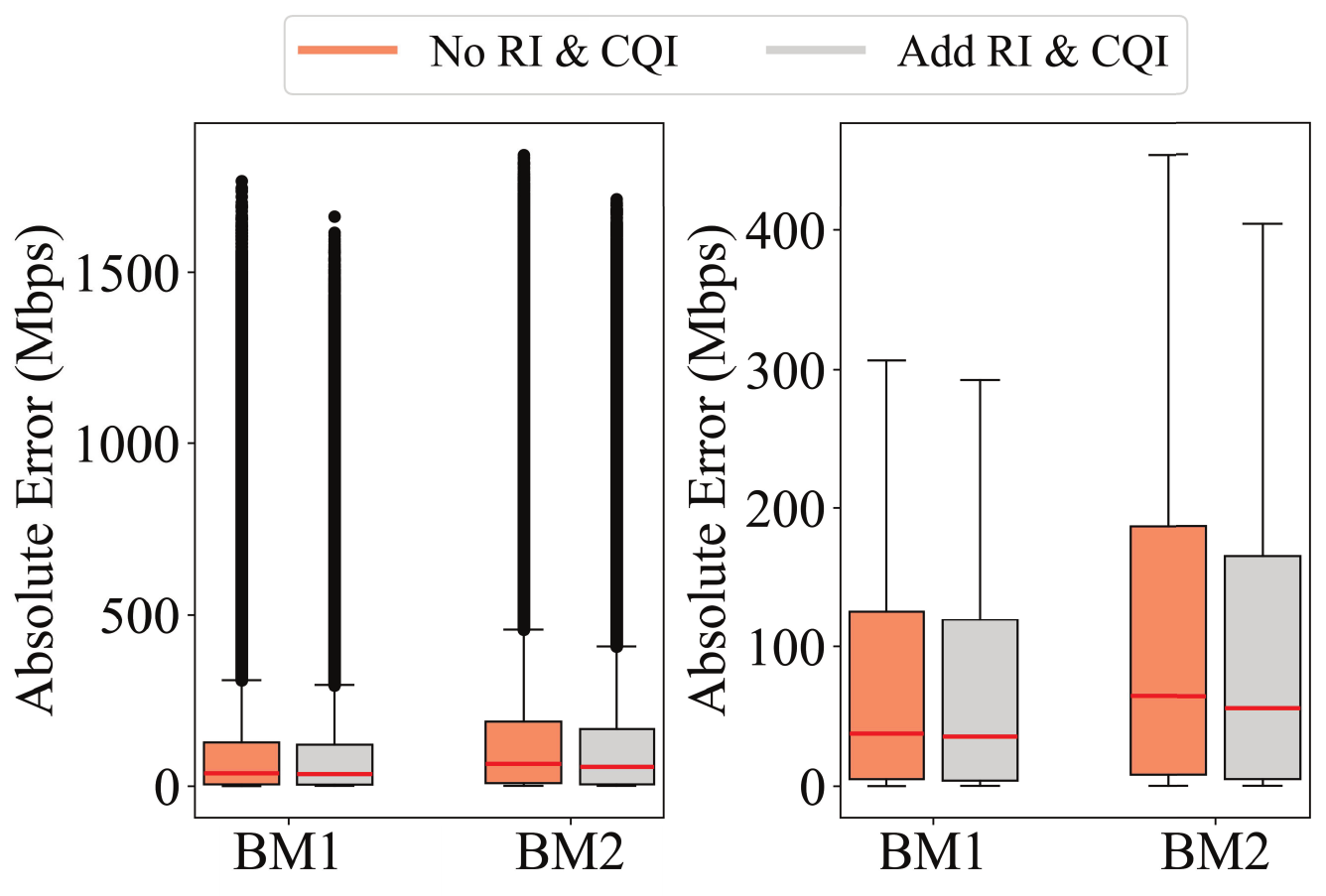} 
    \caption{Absolute error comparison of different input configurations in testing BM1 and testing BM2. The left box plot includes outliers (represented as black dots beyond the whiskers), while the right box plot excludes them. Each box represents the IQR of the absolute error, with the {\rl red} line inside indicating the median value. The whiskers extend to the furthest data points within the range defined as Q1$-1.5\cdot$IQR to Q3$+1.5\cdot$IQR, where Q1 and Q3 denote the first and third quartiles, respectively. Horizontal lines outside the boxes represent the outlier boundaries.}
    \label{fig:Effect of RI and CQI}
\end{figure}

The comparison results, obtained using testing data with 200 sampled sparse points, are presented in Fig.~\ref{fig:Effect of RI and CQI}. The figure illustrates the absolute error distribution between the two U-Net-TP models. The findings clearly indicate that incorporating RI and CQI information significantly reduces the interquartile range (IQR), median, and maximum error magnitude. Specifically, with the 5-channel input configuration, the median absolute error of the predicted Tput Map is 35.37 Mbps and 55.98 Mbps for BM1 and BM2, respectively, while the IQR of the absolute error is 115.37 Mbps and 159.97 Mbps. In contrast, with the 3-channel input configuration, the median absolute error increases to 37.36 Mbps and 64.59 Mbps for BM1 and BM2, respectively, with an IQR of 120.68 Mbps and 178.55 Mbps.
The inclusion of RI and CQI information results in substantial error reductions. This result highlights that incorporating detailed Tput-related features enables U-Net-TP to achieve more accurate predictions, particularly in complex environments such as BM2.

\subsection{Effect of Different Numbers of Sampling Points} \label{Different Sampling Points}

\begin{figure}[tp]
    \centering
    \subfloat[\label{confusion martix 1}]{\includegraphics[width=0.5\linewidth]{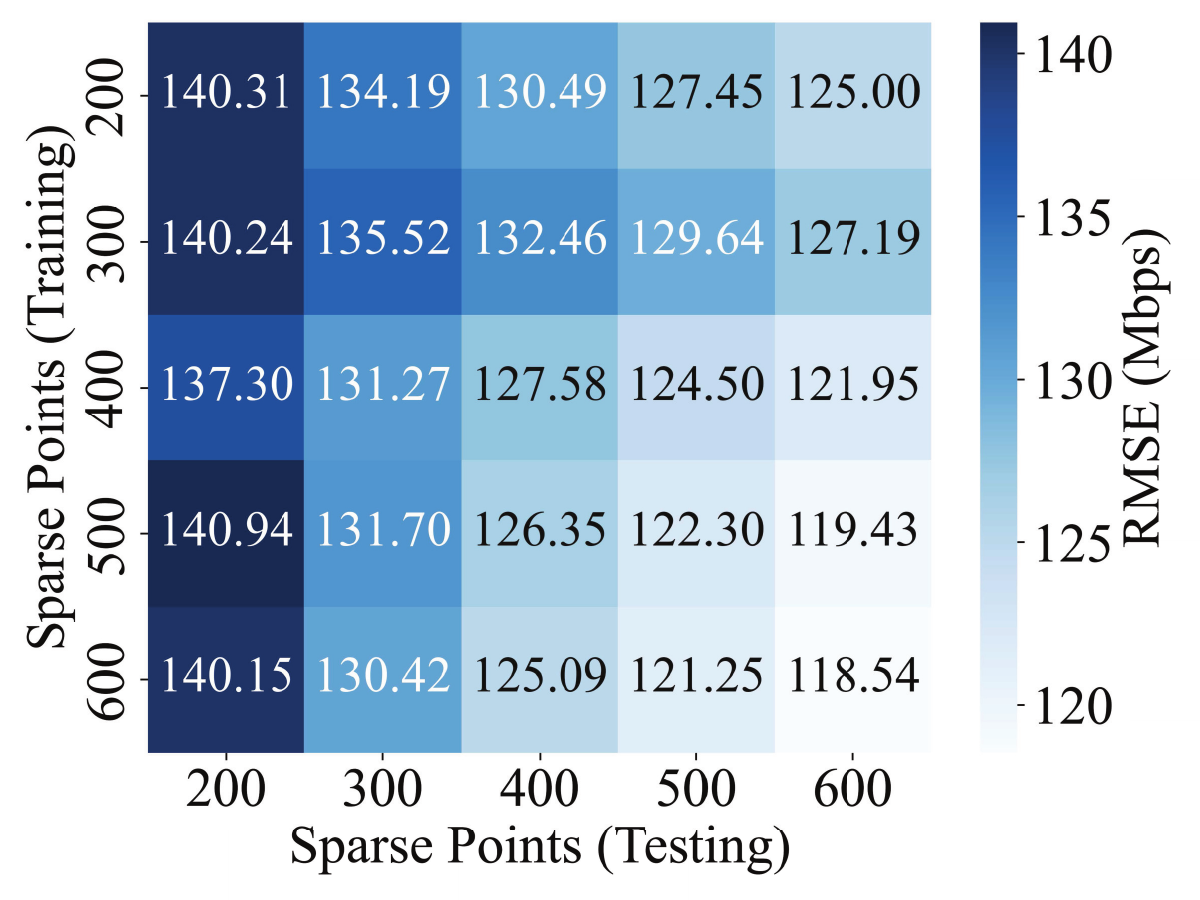}}~
    \subfloat[\label{confusion martix 2}]{\includegraphics[width=0.5\linewidth]{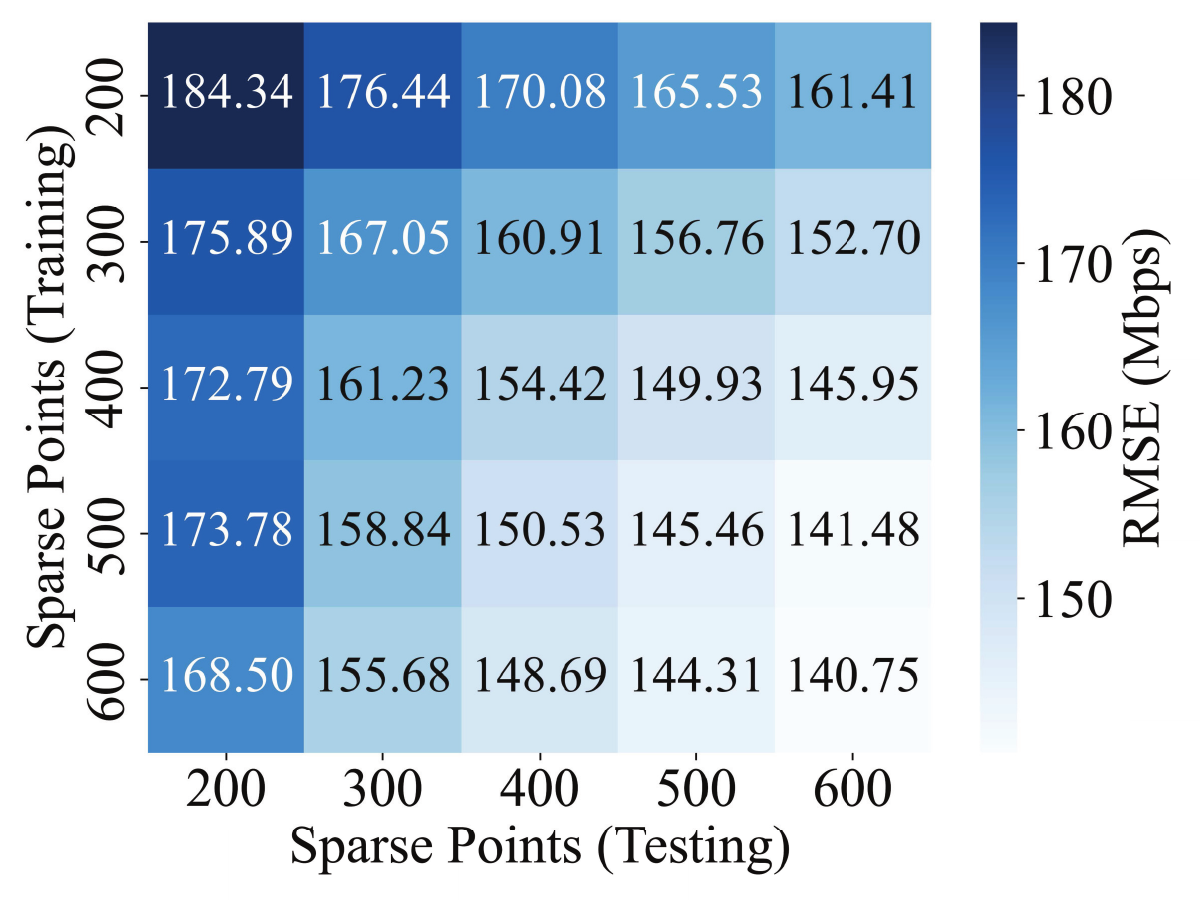}}
    \caption{RMSE values for different combinations of sparse points used in training and testing. (a) RMSE in BM1. (b) RMSE in BM2.}
    \label{fig:Confusion_matrix}
    \vspace{-0.2cm}
\end{figure}

Next, we evaluate the impact of varying the number of sparse sampling points on model performance by training and testing models with different sparse point densities per sparse map.
Fig.~\ref{fig:Confusion_matrix} presents the root MSE (RMSE) for various combinations of sparse points used in training and testing. The results show that increasing the number of sparse points during testing consistently reduces RMSE, regardless of the number of sparse points used in training. This improvement is attributed to the enhanced prediction accuracy afforded by a higher density of sparse points in the test set. Notably, while increasing the number of sparse points during training generally leads to a reduction in RMSE, this trend does not consistently hold across all combinations of training and testing configurations. In certain cases, a higher training point density does not necessarily translate to improved performance. For comparison with conventional approaches, we also present results from linear interpolation and KNN regression using uniform grid samples. Table~\ref{tab:comparison} summarizes the RMSE values, demonstrating that U-Net-TP achieves superior prediction accuracy.

\begin{table}[t]
    \centering
    \caption{RMSE Comparison with 600 Sampled Sparse Points}
    \label{tab:comparison}
    \begin{tabular}{|c|c|c|}
        \hline
        \textbf{Method} & \textbf{BM1 RMSE} & \textbf{BM2 RMSE} \\
        \hline
        Linear Interpolation & 200.35~(Mbps) & 204.06~(Mbps) \\
        KNN Regression       & 162.33~(Mbps) & 177.06~(Mbps) \\
        U-Net-TP             & 118.54~(Mbps) & 140.75~(Mbps) \\
        \hline
    \end{tabular}
    \vspace{-0.3cm}
\end{table}

\begin{figure}[tp]
    \centering
    \includegraphics[width=\linewidth]{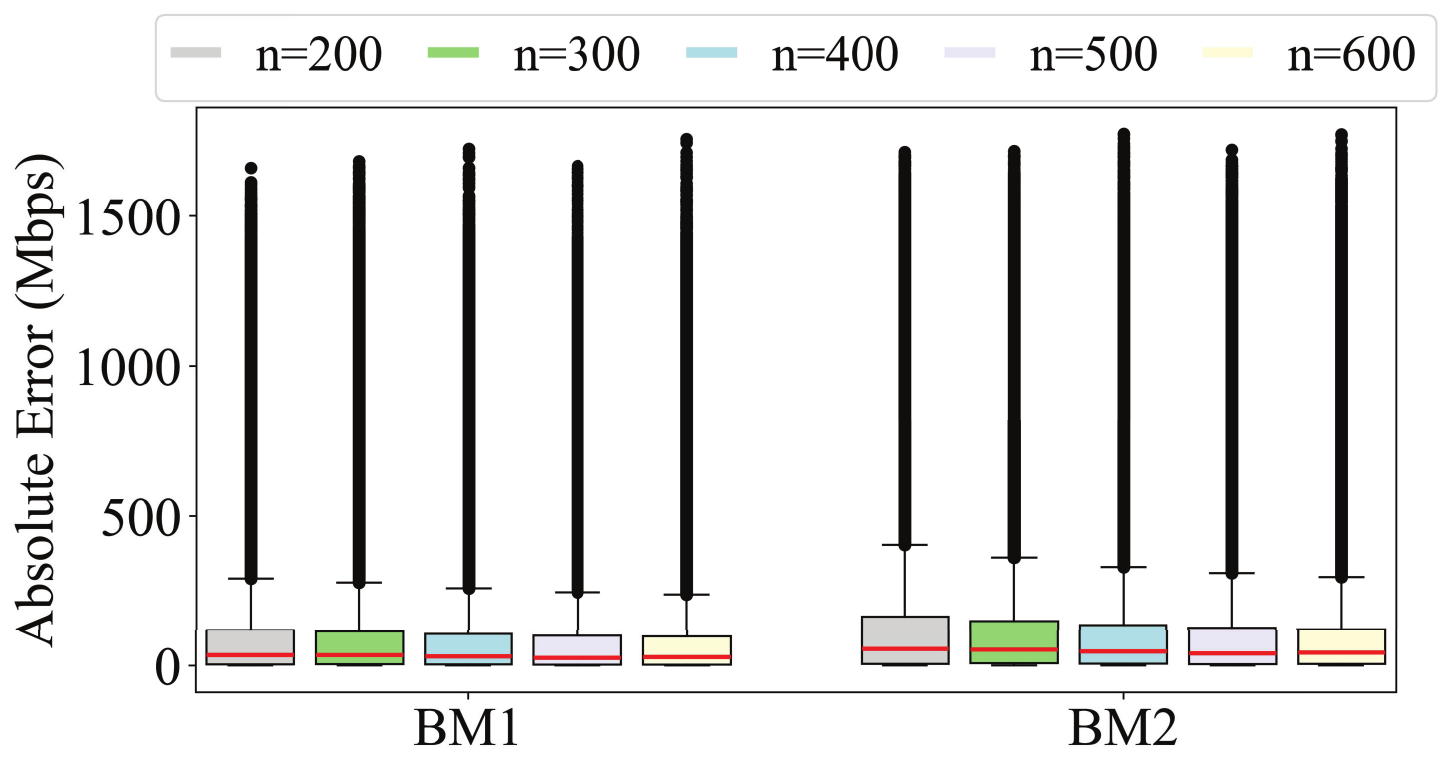}
    \caption{Absolute error comparison for different numbers of sampled sparse points, where $n$ denotes the number of sparse points.}
    \label{fig:Different_sparse_points}
\end{figure}

Fig.~\ref{fig:Different_sparse_points} presents the absolute error for models trained and tested with identical sparse point densities. The results show a consistent reduction in the IQR of absolute errors as the number of sparse points increases, reflecting improved predictive accuracy.
Additionally, the median absolute error decreases with higher sparse point densities, reinforcing the benefits of increased sampling. Specifically, the model achieves median absolute errors of 35.36, 35.65, 31.03, 26.20, and 28.62 Mbps for BM1, and 55.98, 53.98, 47.73, 41.21, and 43.24 Mbps for BM2.
However, the highest errors (outliers) remain largely unchanged despite the increase in sparse points, suggesting that a higher sampling density alone does not effectively mitigate extreme errors. The following section presents a method designed to address these high-error outliers more effectively.

\begin{figure}[t]
    \centering
    \includegraphics[width=\linewidth]{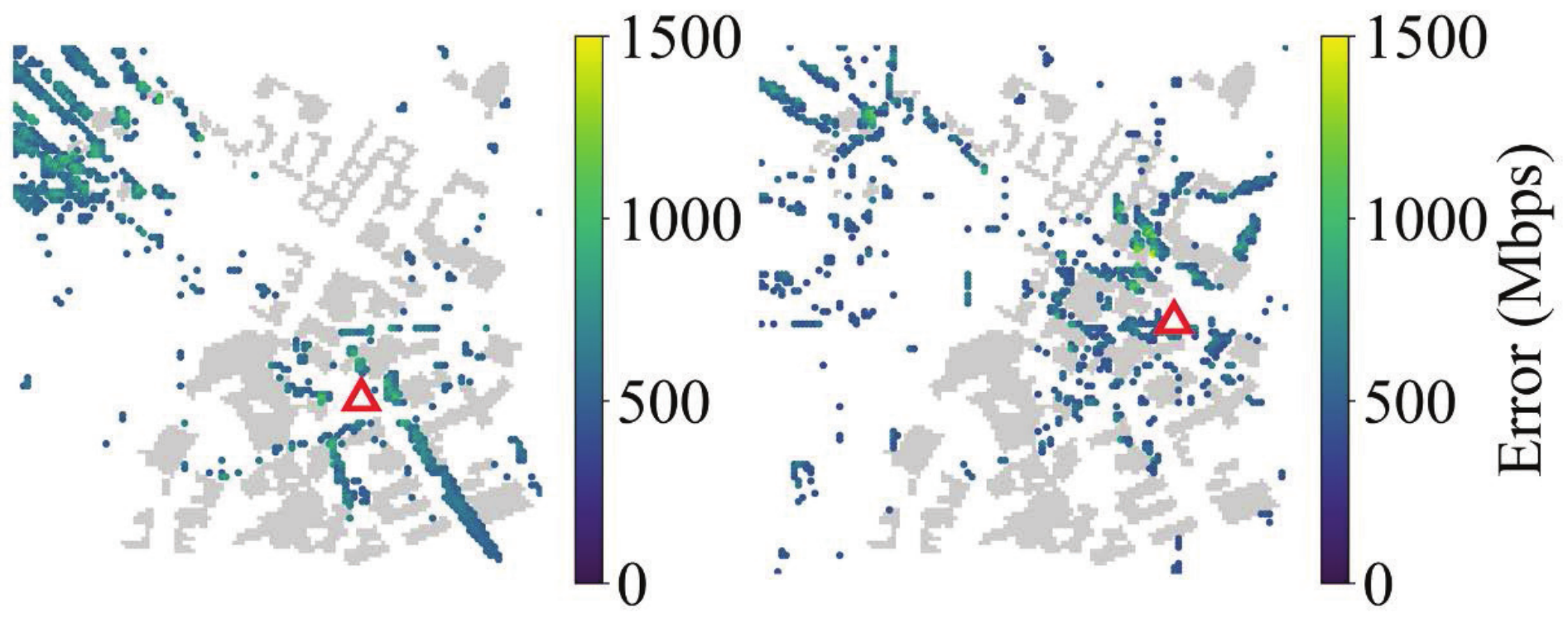}
    \caption{Locations of the top 5\% highest errors in BM1 with 200 sampled sparse points. Error locations are color-coded based on their magnitude, with darker colors indicating higher errors. The red triangle denotes the BS.}
    \label{fig:High_errors}
\end{figure}

\subsection{Dealing with High-Error Outliers}
\label{Dealing with Outliers}

In this subsection, we analyze the causes of high-error outliers and propose a solution to mitigate them. Fig.~\ref{fig:High_errors} highlights the locations of the top 5\% highest errors in BM1, with the left and right plots corresponding to different BS locations. The results show that these high-error regions predominantly occur near building edges or in the spaces between buildings. These errors arise due to sudden Tput variations at building boundaries, especially in areas near the BS where Tput values are higher. Such abrupt transitions challenge the model’s ability to accurately predict Tput, leading to increased errors. This issue is further exacerbated in BM2, where densely clustered buildings create more complex Tput patterns and sharper value changes.

To address high-error regions, we introduce a sparse sampling strategy in which half of the samples are placed at locations with significant throughput variation, and the other half are selected randomly. High-gradient points are identified by computing the gradient of the isotropic PG map, $\qP_{\rm iso}$, using NumPy’s gradient function. The total gradient at each point is computed by summing the absolute gradients along both axes, enabling the selection of regions with large throughput changes. This method, referred to as special sampling, deterministically selects points where prediction errors are more likely to occur.

\begin{figure}[t]
    \centering \includegraphics[width=\linewidth]{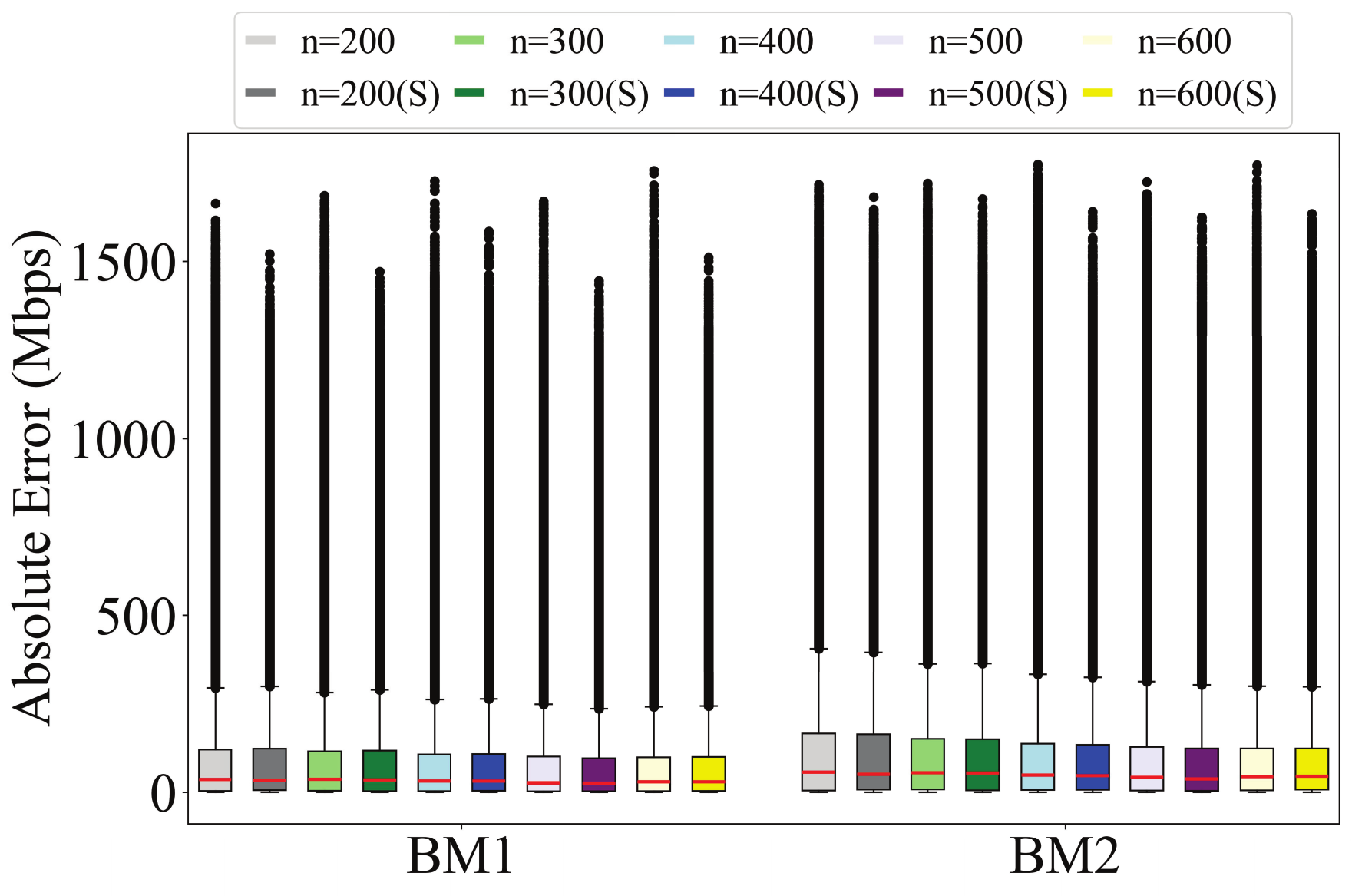}
    \caption{Comparison of absolute errors with and without special sampling in BM1 and BM2. (S) represents special sampling.}
    \label{fig:Special_sampling}
    \vspace{-0.3cm}
\end{figure}

Fig.~\ref{fig:Special_sampling} compares absolute errors before and after applying the special sampling method across five different sampling densities. The results show significant reductions in the highest error values, with improvements of 142.53, 213.91, 142.47, 223.76, and 245.94 Mbps for BM1, and 35.54, 44.02, 135.21, 100.08, and 139.82 Mbps for BM2. These results confirm that the special sampling method effectively mitigates extreme errors and enhances predictive accuracy in both Building Maps.

\subsection{Multi-Output U-Net-TP and Advanced Architectures}
\label{Multiple output U-Net}
The multi-output U-Net-TP architecture (MO U-Net-TP) predicts multiple performance metrics in a single forward pass, offering improved computational efficiency compared to training separate models for each output. Here, we explore a MO U-Net-TP architecture with a 3-channel output image:
$[\mathbf{P}_{\rm dir}, \mathbf{TP}, \mathbf{RI}] \in \mathbb{R}^{N_x \times N_y \times 3} $,
where $\mathbf{P}_{\rm dir}$ represents the directional PG Map, and $\mathbf{TP}$ and $\mathbf{RI}$ correspond to the complete Tput Map and RI Map, respectively. The model is trained and tested using 600 sparse points sampled via the special sampling method.

\begin{figure}[tp]
    \centering
    \subfloat[\label{SOvsMO}]{\includegraphics[width=0.5\linewidth]{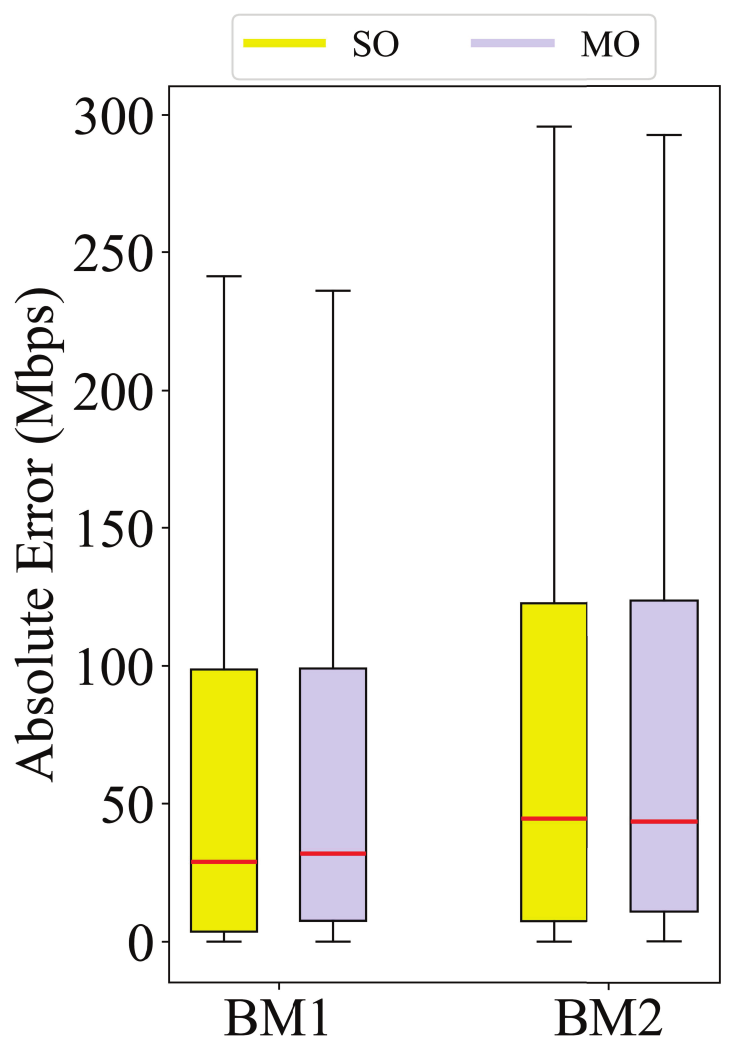}}~
    \subfloat[\label{CDF}]{\includegraphics[width=0.5\linewidth]{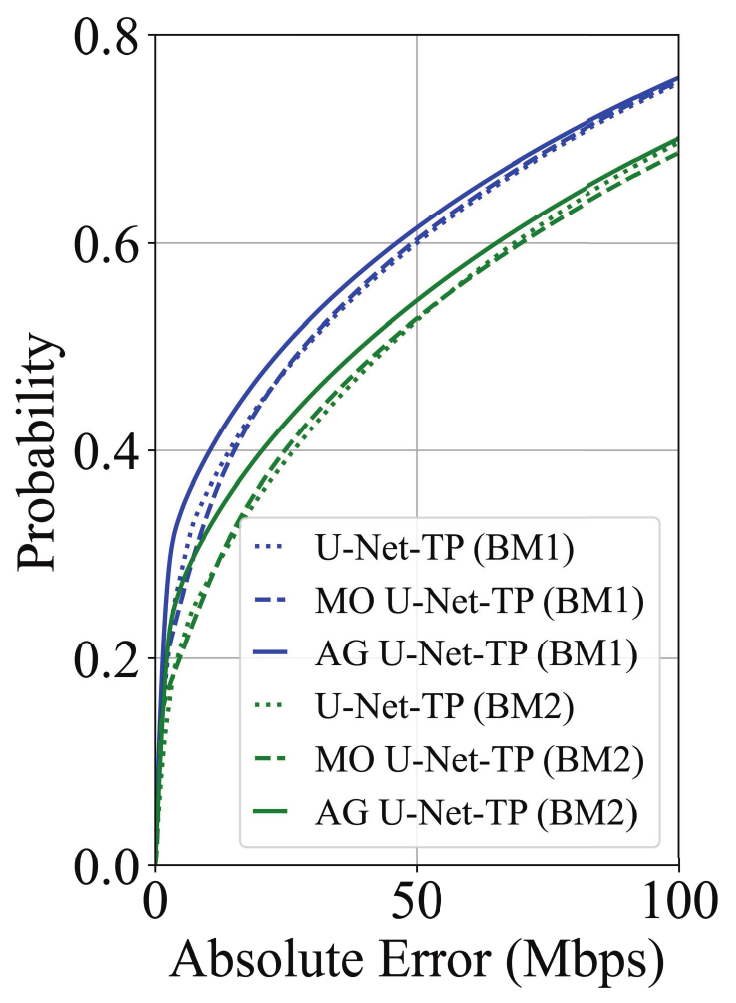}}
    \caption{Absolute error comparison of different U-Net architecture testing in BM1 and BM2. (a) U-Net-TP compared with MO U-Net-TP. (b) CDF of absolute error of U-Net-TP, MO U-Net-TP, and AG U-Net-TP.}
    \label{fig:Diff model comparison}
    \vspace{-0.2cm}
\end{figure}

Fig.~\ref{SOvsMO} compares the performance of U-Net-TP and MO U-Net-TP. The results show that adding additional output channels reduces the IQR. However, predicting multiple performance metrics within a single forward path introduces additional complexity, leading to a slight increase in the median absolute error compared to the single output (SO) configuration.

Although not explicitly depicted in the figures, MO U-Net-TP achieves a median absolute error of 0.15 for the predicted  $\textbf{RI}$ in BM1 and 0.18 in BM2. For $\textbf{P}_{\rm dir}$ prediction, the median absolute error is 2.10 dB for BM1 and 2.18 dB for BM2, which is comparable to the performance of Geo2SigMap \cite{Li-SySPAN24}, a framework specifically designed for PG Map prediction. These results demonstrate that multi-task prediction using the MO framework is a viable and effective approach.

To address the observed increase in median absolute error, we investigate advanced architectural enhancements, such as attention mechanisms. \cite{oktay2018attention} introduced attention gates (AGs), which can be seamlessly integrated into conventional CNNs with minimal computational overhead. AGs leverage coarse-scale contextual information to gate skip connections, allowing only relevant feature activations to be propagated during concatenation, thereby improving the network’s focus on salient regions. We incorporate AGs into the skip connections of the MO U-Net-TP, resulting in the AG U-Net-TP architecture. Fig. \ref{CDF} presents the cumulative distribution function (CDF) of the absolute error for U-Net-TP, MO U-Net-TP, and AG U-Net-TP. The results show that AG U-Net-TP further reduces the median absolute error, achieving 27.35~Mbps for BM1 and 38.62~Mbps for BM2, with an average inference time of 0.23 seconds over 100 trials---representing the best Tput prediction performance among all evaluated configurations.

\section{Conclusion}
We introduced Geo2ComMap, an efficient DL framework for Tput prediction in MIMO-OFDM systems. By leveraging Building Maps, isotropic PG Maps, and approximately 3\% sparse communication measurements, Geo2ComMap accurately reconstructs complete Tput maps and other key communication metrics for a given area. Through extensive experiments, we identified critical input features for the proposed DL model, U-Net-TP, and evaluated its predictive performance. To address the impact of high-error outliers, we introduced a specialized sampling strategy that significantly reduces both extreme errors and RMSE when compared to random sampling. Furthermore, we explored advanced DL architectures: the MO U-Net-TP improves the IQR of absolute errors, while the AG U-Net-TP further reduces overall prediction error. The source code and datasets used in this study are publicly available at~\cite{geo2commap_sourcecode}.


%




\ifCLASSOPTIONcaptionsoff
  \newpage
\fi




\begin{thebibliography}{}
\providecommand{\url}[1]{#1}
\csname url@samestyle\endcsname
\providecommand{\newblock}{\relax}
\providecommand{\bibinfo}[2]{#2}
\providecommand{\BIBentrySTDinterwordspacing}{\spaceskip=0pt\relax}
\providecommand{\BIBentryALTinterwordstretchfactor}{4}
\providecommand{\BIBentryALTinterwordspacing}{\spaceskip=\fontdimen2\font plus
\BIBentryALTinterwordstretchfactor\fontdimen3\font minus
  \fontdimen4\font\relax}
\providecommand{\BIBforeignlanguage}[2]{{%
\expandafter\ifx\csname l@#1\endcsname\relax
\typeout{** WARNING: IEEEtran.bst: No hyphenation pattern has been}%
\typeout{** loaded for the language `#1'. Using the pattern for}%
\typeout{** the default language instead.}%
\else
\language=\csname l@#1\endcsname
\fi
#2}}
\providecommand{\BIBdecl}{\relax}
\BIBdecl

\end{thebibliography}


\begin{thebibliography}{10}
\providecommand{\url}[1]{#1}
\csname url@samestyle\endcsname
\providecommand{\newblock}{\relax}
\providecommand{\bibinfo}[2]{#2}
\providecommand{\BIBentrySTDinterwordspacing}{\spaceskip=0pt\relax}
\providecommand{\BIBentryALTinterwordstretchfactor}{4}
\providecommand{\BIBentryALTinterwordspacing}{\spaceskip=\fontdimen2\font plus
\BIBentryALTinterwordstretchfactor\fontdimen3\font minus
  \fontdimen4\font\relax}
\providecommand{\BIBforeignlanguage}[2]{{%
\expandafter\ifx\csname l@#1\endcsname\relax
\typeout{** WARNING: IEEEtran.bst: No hyphenation pattern has been}%
\typeout{** loaded for the language `#1'. Using the pattern for}%
\typeout{** the default language instead.}%
\else
\language=\csname l@#1\endcsname
\fi
#2}}
\providecommand{\BIBdecl}{\relax}
\BIBdecl

\bibitem{Zhang-24ComMag}
S.~Zhang \emph{et~al.}, ``Physics-inspired machine learning for radiomap
  estimation: Integration of radio propagation models and artificial
  intelligence,'' \emph{{IEEE} Commun. Mag.}, vol.~62, no.~8, pp. 155--161,
  2024.

\bibitem{Zhou-25ArXiv}
Y.~Zhou \emph{et~al.}, ``Efficient transmission of radiomaps via
  physics-enhanced semantic communications,'' \emph{arXiv preprint
  arXiv:2501.10654}.

\bibitem{elsayed2024ofdm}
E.~Elsayed, ``Investigations on ofdm uav-based free-space optical transmission
  system with scintillation mitigation for optical wireless
  communication-to-ground links in atmospheric turbulence,'' \emph{Opt Quant
  Electron}, vol.~56, p. 837, 2024.

\bibitem{Levie-21TW}
R.~Levie \emph{et~al.}, ``{RadioUNet}: Fast radio map estimation with
  convolutional neural networks,'' \emph{{IEEE} Trans. Wireless Commun.},
  vol.~20, no.~6, pp. 4001--4015, 2021.

\bibitem{Li-SySPAN24}
Y.~Li \emph{et~al.}, ``{Geo2SigMap}: High-fidelity {RF} signal mapping using
  geographic databases,'' in \emph{2024 IEEE Int. Symp. Dynamic Spectrum Access
  Networks (DySPAN)}, 2024, pp. 277--285.

\bibitem{Yapar-24}
{\c{C}}.~Yapar \emph{et~al.}, ``Overview of the first pathloss radio map
  prediction challenge,'' \emph{{IEEE} Open J. Signal Process.}, vol.~5, pp.
  948--963, 2024.

\bibitem{3gpp_38_306}
{3GPP}, ``{User Equipment (UE) radio access capabilities},'' 3GPP, Tech. Rep.
  TS 38.306 V16.3.0 (2021-01), 2021.

\bibitem{openstreetmap}
\BIBentryALTinterwordspacing
``{OpenStreetMap (OSM)},'' {2023}. [Online]. Available:
  \url{https://www.openstreetmap.org/}
\BIBentrySTDinterwordspacing

\bibitem{blender}
\BIBentryALTinterwordspacing
``{Blender},'' {2023}. [Online]. Available: \url{https://www.blender.org/}
\BIBentrySTDinterwordspacing

\bibitem{sionna_rt}
J.~Hoydis \emph{et~al.}, ``{Sionna RT}: Differentiable ray tracing for radio
  propagation modeling,'' \emph{arXiv preprint arXiv:2303.11103}.

\bibitem{3gpp_38_214}
{3GPP}, ``{3GPP TS 38.214},'' 3GPP, Tech. Rep. TS 38.214 V17.1.0 (2022-04),
  2022.

\bibitem{oktay2018attention}
O.~Oktay \emph{et~al.}, ``Attention {U-Net}: Learning where to look for the
  pancreas,'' \emph{arXiv preprint arXiv:1804.03999}.

\bibitem{geo2commap_sourcecode}
\BIBentryALTinterwordspacing
``{Geo2ComMap source code},'' {2025}. [Online]. Available:
  \url{https://github.com/geo2commap/Geo2ComMap}
\BIBentrySTDinterwordspacing

\end{thebibliography}
%


\sloppy
\bibliographystyle{IEEEtran}

\fussy

%








\end{document}